\documentclass[sigconf,screen]{acmart}

\usepackage{hyperref}
\usepackage{booktabs}
\usepackage{multirow}
\usepackage{enumerate}
\usepackage{subcaption}
\usepackage{movie15}

\AtBeginDocument{%
  }

\newcommand{\FigDigit}[1]{
  \begin{subfigure}{0.11\textwidth}
  \includemovie[poster=images/samples/#1.png,mouse,repeat]{1\linewidth}{.7364\linewidth}{audios/#1.mp3}
  \label{fig:#1}
\end{subfigure}}

\copyrightyear{2024}
\acmYear{2024}
\setcopyright{acmlicensed}
\acmConference[ICMR '24]{Proceedings of the 2024 International Conference on Multimedia Retrieval}{June 10--14, 2024}{Phuket, Thailand}
\acmBooktitle{Proceedings of the 2024 International Conference on Multimedia Retrieval (ICMR '24), June 10--14, 2024, Phuket, Thailand}
\acmPrice{15.00}
\acmDOI{10.1145/3652583.3658086}
\acmISBN{979-8-4007-0619-6/24/06}

\begin{document}

\title{Retrieval-Augmented Audio Deepfake Detection}


\author{Zuheng Kang}
\orcid{0000-0001-9789-7798}
\authornote{Both authors contributed equally to this research}
\affiliation{%
  \institution{Ping An Technology (Shenzhen) Co., Ltd.}
  \streetaddress{Jialian Payment Building, Nanshan District}
  \city{Shenzhen}
  \postcode{518063}
  \country{China}}
\email{kangzuheng896@pingan.com.cn}

\author{Yayun He}
\authornotemark[1]
\orcid{0009-0008-9722-8031}
\affiliation{%
  \institution{Ping An Technology (Shenzhen) Co., Ltd.}
  \streetaddress{Jialian Payment Building, Nanshan District}
  \city{Shenzhen}
  \postcode{518063}
  \country{China}}
\email{heyayun097@pingan.com.cn}

\author{Botao Zhao}
\orcid{0000-0001-9114-1236}
\affiliation{%
  \institution{Ping An Technology (Shenzhen) Co., Ltd.}
  \streetaddress{Jialian Payment Building, Nanshan District}
  \city{Shenzhen}
  \postcode{518063}
  \country{China}}
\email{zhaobotao204@pingan.com.cn}

\author{Xiaoyang Qu}
\orcid{0000-0001-8353-4064}
\affiliation{%
  \institution{Ping An Technology (Shenzhen) Co., Ltd.}
  \streetaddress{Jialian Payment Building, Nanshan District}
  \city{Shenzhen}
  \postcode{518063}
  \country{China}}
\email{quxiaoyang343@pingan.com.cn}

\author{Junqing Peng}
\orcid{0009-0009-2903-9615}
\affiliation{%
  \institution{Ping An Technology (Shenzhen) Co., Ltd.}
  \streetaddress{Jialian Payment Building, Nanshan District}
  \city{Shenzhen}
  \postcode{518063}
  \country{China}}
\email{pengjq@pingan.com.cn}

\author{Jing Xiao}
\orcid{0000-0001-9615-4749}
\affiliation{%
  \institution{Ping An Insurance (Group) Company of China}
  \streetaddress{Ping An Finance Center (PAFC), Futian District}
  \city{Shenzhen}
  \postcode{518047}
  \country{China}}
\email{xiaojing661@pingan.com.cn}

\author{Jianzong Wang}
\orcid{0000-0002-9237-4231}
\authornote{Corresponding author: Jianzong Wang, jzwang@188.com}
\affiliation{%
  \institution{Ping An Technology (Shenzhen) Co., Ltd.}
  \streetaddress{Jialian Payment Building, Nanshan District}
  \city{Shenzhen}
  \postcode{518063}
  \country{China}}
\email{jzwang@188.com}

\renewcommand{\shortauthors}{Zuheng Kang et al.}

\begin{abstract}

  With recent advances in speech synthesis including text-to-speech (TTS) and voice conversion (VC) systems enabling the generation of ultra-realistic audio deepfakes, there is growing concern about their potential misuse.
  However, most deepfake (DF) detection methods rely solely on the fuzzy knowledge learned by a single model, resulting in performance bottlenecks and transparency issues.
  Inspired by retrieval-augmented generation (RAG), we propose a retrieval-augmented detection (RAD) framework that augments test samples with similar retrieved samples for enhanced detection.
  We also extend the multi-fusion attentive classifier to integrate it with our proposed RAD framework.
  Extensive experiments show the superior performance of the proposed RAD framework over baseline methods, achieving state-of-the-art results on the ASVspoof 2021 DF set and competitive results on the 2019 and 2021 LA sets.
  Further sample analysis indicates that the retriever consistently retrieves samples mostly from the same speaker with acoustic characteristics highly consistent with the query audio, thereby improving detection performance.

\end{abstract}




\maketitle


\section{Introduction}

Recent artificial intelligence (AI) techniques have enabled the generation of synthesized audio known as DeepFakes (DF) with increasing degrees of fidelity to natural human speech.
Sophisticated DF generation techniques, such as text-to-speech (TTS) and voice conversion (VC) to mimic the timbre, prosody, and intonation of the speaker, can now generate audio perceptually indistinguishable from genuine recordings.
However, the potential malicious use of such AI-synthesized speech has serious personal and societal implications, including disruption of automatic speaker verification (ASV) systems, propagating misinformation, and defaming reputations.
As a result, the vulnerability of current audio-based communication systems to synthesized speech poses a serious threat.
Therefore, the development of effective DF detection techniques is urgently needed to address this emerging risk for everyone of us.

Recent advances in artificial intelligence generation content (AIGC) techniques, such as DF generation, however, have made detecting DFs increasingly challenging.
In newly organized competitions, such as the ASVspoof 2021 \cite{Yamagishi2021ASVspoof2A}, ADD 2022 \cite{yi2022add}, 2023 \cite{Yi2023ADD2T}, even the state-of-the-art (SOTA) DF detection systems perform poorly, with an equal error rate (EER) of over 10\% \cite{Yi2023AudioDD}, making them impossible to detect properly, and unsuitable for commercial deployment.
This suggests that the rapid development of DF generation seems to have greatly outpaced the development of detection technologies -- the traditional detection methods are far from inadequate for identifying current out-of-domain \cite{chen2024gaia} DF-generated samples.
Building robust and reliable DF detection systems remains a critical issue for the research community that has not yet been fully resolved.

Recent decades have witnessed the emergence of various frameworks for detecting DF audio.
The predominant frameworks utilize a pipeline consisting of a front-end feature extractor and a back-end classifier (discussed in $ \S $ \ref{sec:traditional}).
Early works relied on hand-crafted features for DF detection with some success, such as Mel-frequency
cepstral coefficients (MFCC) \cite{hamza2022deepfake}, linear-frequency cepstral
coefficients (LFCC) \cite{Davis1980ComparisonOP,wang2021comparative,luo2021capsule}, CQT \cite{ziabary2021countermeasure,li2021channel}, and F0 sub-band feature \cite{Fan2023SpatialRL}.
However, these features exhibit limited performance due to the limited dataset used to train the robust model.
More recent SOTA frameworks have leveraged the capabilities of self-supervised model feature extractors, such as wav2vec \cite{Baevski2020wav2vec2A,MartinDonas2022TheVA,Wang2021InvestigatingSF,Tak2022AutomaticSV,cai2024integrating} and WavLM \cite{Chen2021WavLMLS,Guo2023AudioDD,cai2024integrating}.
Moreover, Kawa et al. \cite{Kawa2023ImprovedDD} utilize another powerful task-specific feature in deepfake detection using the pre-trained Whisper model \cite{Radford2022RobustSR}.
Theoretically, by training on a large number of labeled or unlabeled bonafide samples in very large datasets, these feature extractor models could be particularly sensitive to unseen DF artifacts.
Some alternative frameworks replace hand-crafted features with end-to-end trainable encoders.
Notable examples include Jung et al.'s advanced graph attention network architecture AASIST \cite{Jung2021AASISTAA}, and Huang et al.'s discriminative frequency-based improvements to SincNet \cite{Huang2023DiscriminativeFI}, both of which achieve competitive performance.
As discussed by Sun et al. \cite{Sun2023AISynthesizedVD}, vocoders employed in regular speech synthesis can introduce inconsistencies that reveal DFs.
However, regardless of the methodology and architecture used, these frameworks rely solely on a single model to accomplish this challenging task, which may prove inadequate.

DF detection should be a knowledge-intensive task that also relies on vast external knowledge.
To explain this, let's start with a story.
In the identification of antique artifacts, fakes are often fabricated so realistically that it is difficult to determine their authenticity.
Senior experts usually conduct a meticulous comparative analysis that includes material and textural attributes of many similar artifacts.
With so many subtle factors to consider, relying solely on one's limited intelligence is likely to result in poor judgment.
Similarly, relying only on a single model to detect deepfakes may be too challenging to make mistakes.

Retrieval augmented generation (RAG) \cite{Lewis2020RetrievalAugmentedGF,Gao2023RetrievalAugmentedGF} methodology provides a nice example for solving the problem of knowledge-intensive tasks (discussed in $ \S $ \ref{sec:rag}).
The RAG resolved the limitation of the single model by combining a pre-trained large language model (LLM) with an information retrieval system over a large knowledge database.
Fundamentally, the RAG framework leverages significant supplementary real-time updated additional interpretable knowledge (with rapidly changing, proprietary data sources) to augment the limitations of a single model's knowledge, enabling it to provide a more reasoned answer.
Complementary background information allows a single model to overcome its inherent knowledge gaps.

Similarly, the retrieval-augmented-based approach could provide the same benefits for DF detection.
When analyzing suspect audio, the model could query a retrieval system to find many similar reference audio segments.
These retrieved results would provide additional references to inform the deepfake detection, such as typical artifacts for bonafide or synthetic spoofed examples.
Then, the deepfake detector could integrate these retrieval results and the suspected audio into its decision process.
This retrieval-augmented approach has several advantages.
The model gains access to a much larger knowledge base beyond what can be encoded only in its model parameters.
The retrieved results also provide supporting evidence for decisions, improving model performance.
Furthermore, the system can update or modify its knowledge database for different detection tasks, since the model has learned to characterize a limited type of DF synthesizing methods.
In summary, augmenting deepfake detectors with conditional retrieval of external data is an attractive direction.
Developing DF detection methodology with a retrieval-augmented approach is worthy of further exploration.
To implement the above-mentioned issue, we made the following contributions:

\begin{itemize}
  \item We proposed a retrieval augmented detection (RAD) framework which innovatively retrieves similar samples with additional knowledge, and incorporates them into the detection process for improved detection performance.
  \item We extend the multi-fusion attentive classifier to integrate with RAD.
  \item Extensive Experiments show that our proposed method achieves state-of-the-art results on the ASVspoof 2021 DF set and competitive results on the 2019 and 2021 LA sets, demonstrating the effectiveness of the proposed retrieval-augmented approaches.
\end{itemize}

\section{Preliminary}

\subsection{Traditional Frameworks}
\label{sec:traditional}


\subsubsection*{(1) Pipeline Framework}
The most common pipelines typically include separate components for feature extraction, and classification (Figure \ref{fig:frameworks}-1).
Specifically, the front-end feature extractor first converts the raw speech signal $ x $ into a speech features $ y $.
These speech features are then passed to the back-end classifier, which analyzes the speech features that make a bonafide vs. spoof decision $ z $.
Such architectures leverage the capabilities of efficient hand-crafted features or semantically rich self-supervised pre-trained features to obtain highly informative speech representations.
The back-end classifier then fully analyzes and mines these features, and makes the final prediction.

\subsubsection*{(2) End-to-End Framework}
More advanced frameworks transform the feature extraction into a trainable encoder, forming an end-to-end architecture (Figure \ref{fig:frameworks}-2).
Differently, instead of the feature extractor, the raw speech $ x $ is fed into a trainable encoder to produce speech representations $ y $.
To further improve performance, some approaches fuse multiple feature extractors or encoders, concatenating the resulting speech features and representations for deeper training.

\subsection{Retrieval Augmented Generation}
\label{sec:rag}

Prior to introducing our proposed approach, it is instructive to first provide background knowledge on the retrieval augmented generation (RAG) framework \cite{Lewis2020RetrievalAugmentedGF} to facilitate comparison with our method.
The RAG framework consists of three main stages:

\begin{figure}[t]
  \includegraphics[width=0.48\textwidth]{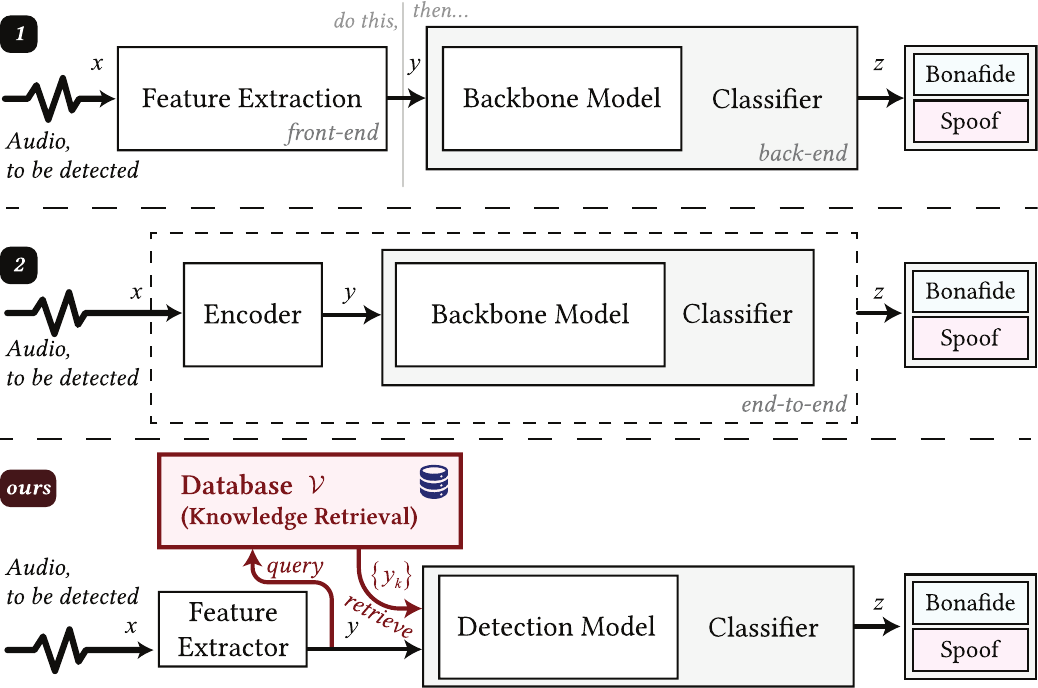}
  \centering
  \caption{The overview of traditional frameworks, and our proposed framework for audio deepfake detection.
    (1) shows the pipeline framework.
    (2) shows the end-to-end framework.
    (ours) shows our proposed retrieval augmented-based detection (RAD) framework.}
  \label{fig:frameworks}
\end{figure}

\subsubsection*{(1) Build Knowledge Retrieval Database}

As shown in the stage 1 (blue section) of Figure \ref{fig:ragd}-RAG, the plain text and format-rich text database $ x $ is partitioned into smaller chunks $ \left\{x_n\right\} $ at $ n^\mathrm{th} $ sample.
These text chunks are then embedded into dense vector representations $ \left\{v_n\right\} $ by a language model.
In addition, each embedding $ v_n $ maintains an index that links it to its original text chunk $ x_n $, allowing the retrieval of the original text content.
Finally, these embeddings $ \left\{v_n\right\} $ can be stored in a vector database $ \mathcal{V} $ that facilitates efficient similarity search and retrieval.

\subsubsection*{(2) Retrieve Knowledge}

As shown in the stage 2 (red section) of Figure \ref{fig:ragd}-RAG, the user's query text $ \tilde{x}_q $ is embedded into a query embedding $ \tilde{v}_q $ using the same language model.
This query embedding is then leveraged to perform a similarity search across the vector database $ \mathcal{V} $ containing all the document chunk embeddings.
The top $ K $ most similar embeddings $ \left\{\tilde{v}_k\right\} $ related to its document chunks $ \left\{\tilde{x}_k\right\} $ are retrieved based on their semantic proximity to the query embedding $ \tilde{v}_q $ in the vector space.
These most relevant $ K $ document chunks $ \left\{\tilde{x}_k\right\} $ can be used as augmented contextual information to complement the original user query.

\subsubsection*{(3) Get Results (Answer Generation)}

As shown in stage 3 (green section) of Figure \ref{fig:ragd}-RAG, the original user query $ \tilde{x}_q $ is concatenated with the retrieved document chunks $ \left\{\tilde{x}_k\right\} $ to construct an expanded prompt $ p $ with its function $ \mathcal{P} $.
Where $ p=\mathcal{P} \left( \tilde{x}_q,\left\{ \tilde{x}_k \right\} \right) $.
This enriched prompt $ p $, which contains both the initial query and relevant contextual information, is subsequently input to a large language model (LLM).
The LLM analyzes the overall content and relationships within $ p $ to generate the final answer $ z $.

\section{Methodology}

\subsection{Self-supervised Feature with WavLM}
\label{sec:wavlm}

The overall framework of our proposed method is illustrated in Figure \ref{fig:frameworks}.
Unlike traditional methods (described in $ \S $ \ref{sec:traditional}), our proposed framework leverages the state-of-the-art WavLM \cite{Chen2021WavLMLS} feature extractor and incorporates an additional retrieval module after feature extraction to overcome performance bottlenecks.
Specifically, we adopt a retrieval-augmented structure similar to RAG (described in $ \S $ \ref{sec:rag}), which retrieves a few similar features from the bonafide samples and fuse them with the original test features before feeding into the detection model.
By incorporating retrieved features highly similar to the test sample, our model can make much more reliable predictions through joint analysis.

In the following section, we describe these modules in detail, including the WavLM feature extractor (described in $ \S $ \ref{sec:wavlm}), the retrieval augmented mechanism (described in $ \S $ \ref{sec:rad}), and the design of the detection model classifier (described in $ \S $ \ref{sec:detect}).
Meanwhile, since this framework is complex, we present a number of speed-up techniques (described in $ \S $ \ref{sec:performance}) to greatly reduce the space and time complexity.
To further improve performance, we also jointly optimize the WavLM feature extractor with the detection model in an end-to-end manner (described in $ \S $ \ref{sec:finetune}).

\begin{figure}[b]
  \includegraphics[width=0.475\textwidth]{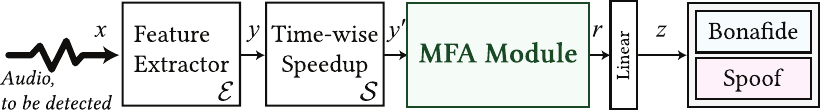}
  \centering
  \caption{The baseline structure for fine-tuning.}
  \label{fig:baseline}
\end{figure}

\subsubsection*{WavLM Feature Extraction}
Recent advanced feature extractor WavLM \cite{Chen2021WavLMLS} employs wav2vec 2.0 \cite{Baevski2020wav2vec2A} as its backbone and is trained with larger real multilingual, multi-channel unlabeled speech data for much better performance.
WavLM utilizes a masked speech denoising and prediction framework that artificially adds noise and overlapping speech to clean input audio before masking certain time segments, and the model should then predict the speech content of the original frames of these masked segments.
This denoising process allows WavLM to learn robust representations that capture not only a variety of speech features but also the acoustic environment.
In addition, WavLM performs excellently in a variety of downstream speech tasks such as automatic speech recognition (ASR), automatic speaker verification (ASV), and text-to-speech (TTS) with minimal fine-tuning.
This suggests that WavLM already understands and is familiar with many high-level speech characteristics of bonafide audio, which are particularly appropriate for unseen DF-synthesized audio, since it often contains features that are very different from bonafide audio.

In the proposed framework, the feature extraction component utilizes the complete set of latent features $ y \in \mathbb{R} ^{L\times T\times F} $ from all layers of the WavLM encoder transformer when processing an input audio segment $ x $.
Where $ L $ is the number of WavLM encoder transformer layers, $ T $ is the number of frames, $ F $ is the dimension of features (same as WavLM feature size).
This enables the model to leverage speech information encompassing low-level acoustic features as well as higher-level semantic abstractions extracted by the deeper layers.

\begin{figure*}[t]
  \includegraphics[width=0.999\textwidth]{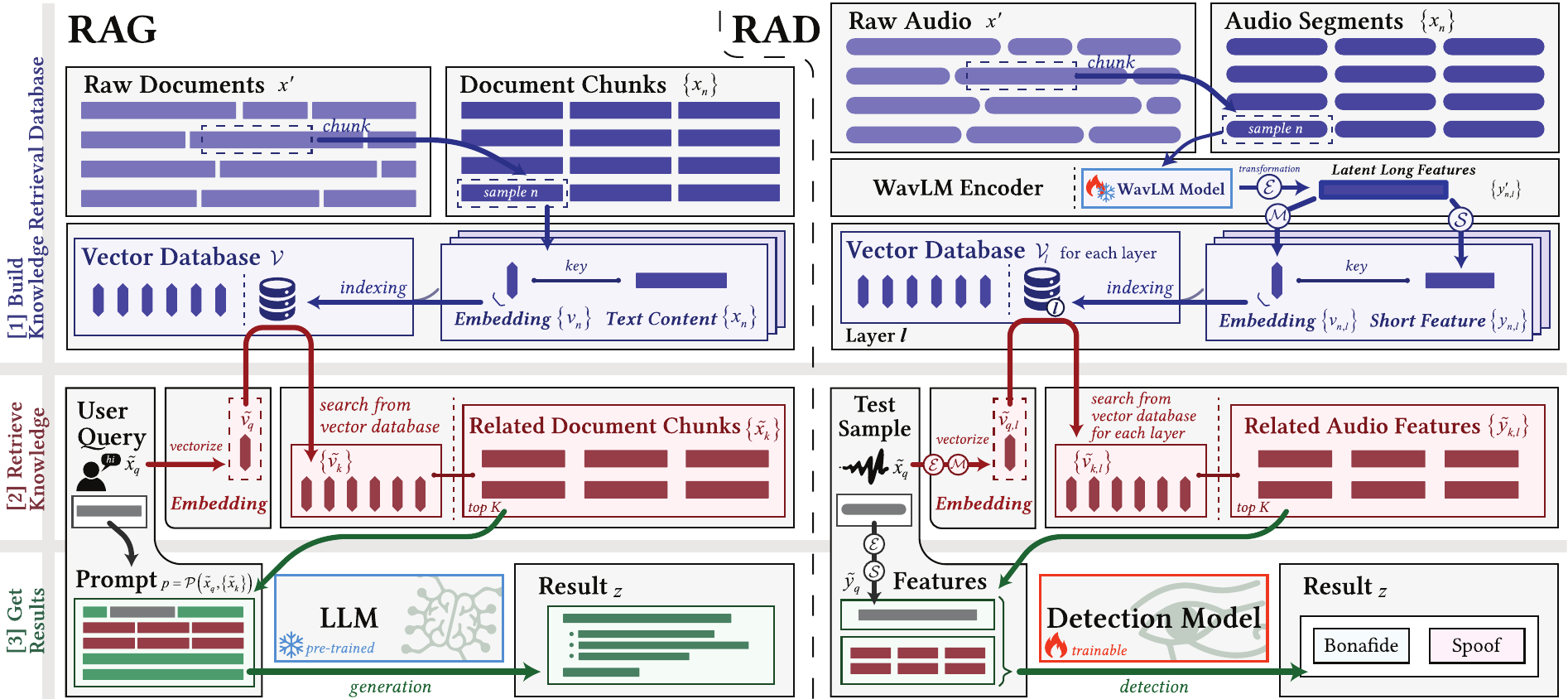}
  \centering
  \caption{The overview of the RAG and RAD pipeline.
    Triangular edge rectangles represent vectors for retrieval databases.
    In RAG, long rectangles represent document chunks.
    In RAD, long rectangles with/without an outline represent long/short features,
    rounded edge rectangles represent audio segments.}
  \label{fig:ragd}
\end{figure*}

\subsubsection*{WavLM Fine-Tuning}
\label{sec:finetune}

Since WavLM is trained only on bonafide audio during pre-training, it may not have exposure to spoofed samples, which potentially leads to incorrect classifications.
Therefore, we first fine-tune the entire WavLM feature extractor in an end-to-end manner without the RAD framework.
That is, shown in Figure \ref{fig:baseline}, the speech is encoded into short features by a trainable WavLM model $ \mathcal{E} $ and time-wise speedup method $ \mathcal{S} $, which is then encoded into intermediate representations by an MFA module.
These representations are then classified as either bonafide or spoofed by a fully connected layer.
By jointly optimizing the parameters, we obtain a fine-tuned WavLM model that can serve as an improved feature extractor in the subsequent RAD framework.
In the subsequent RAD inference phase, we only leverage the fine-tuned WavLM and discard the back-end model.

\subsection{Retrieval Augmented Detection}
\label{sec:rad}

To address the performance limitations imposed by the detection bottleneck, we propose the retrieval augmented detection (RAD) framework.
Similar to the RAG framework, the proposed RAD approach consists of three main stages, but with some procedural modifications compared to RAG.

\subsubsection*{(1) Build Knowledge Retrieval Database}

As shown in the stage 1 (blue section) of Figure \ref{fig:ragd}-RAD, the bonafide audio dataset $ x^\prime $ is segmented into smaller audio segments $ \left\{x_n\right\} $ at the $ n^\mathrm{th} $ sample.
These audio segments can be encoded to latent long feature representations $ \left\{ y^\prime_{n,l} \right\} \in \mathbb{R} ^{N\times L\times T^\prime \times F} $ by WavLM feature extractor $ \mathcal{E}\left(\cdot\right) $, where $ T^\prime $ is the time dimension of long features, $ N $ is the number of audio segments, and $ l $ indexes the encoder layer of WavLM.
Subsequently, two operations are performed on the long features $ \left\{ y^\prime_{n,l} \right\} $:
(a) The features are embedded into dense vector representations $ \left\{v_{n,l} \right\} \in \mathbb{R} ^{N\times L\times F} $ via the mapping $ \mathcal{M}\left(\cdot\right) $ in Equation \ref{eq:M}, where each $ v_{n,l} $ summarizes the time-wise features $ y^\prime _{n,l} $ by temporal averaging to eliminate the time dimension $ T^\prime $;
(b) The time dimension is shortened to form short feature $ \left\{ y_{n,l} \right\} \in \mathbb{R} ^{N\times L\times T \times F} $ for improved efficiency by the function $ \mathcal{S}\left(\cdot\right) $ (details in $ \S $ \ref{sec:speedup}), where $ T $ is the time dimension of short feature.
\begin{equation}
  v_{n,l}=\mathcal{M} \left( y^\prime_{n,l} \right) =\frac{1}{T}\sum_{t=1}^T{\left. y^\prime_{n,l} \right|_t}.
  \label{eq:M}
\end{equation}
Importantly, each embedding $ v_{n,l} $ maintains an index linking to its original short feature $ y_{n,l} $, enabling the retrieval of the original audio segment $x_n$ for the source content.
Finally, the collection of embeddings $ \left\{v_{n,l} \right\} $ are stored in $ l $ vector databases $ \mathcal{V}_l $ to enable efficient similarity search and retrieval.

\subsubsection*{(2) Retrieve Knowledge}

As shown in the stage 2 (red section) of Figure \ref{fig:ragd}-RAD, a sample to be detected $ \tilde{x}_q $ can be embedded into a query embedding $ \tilde{v}_{q,l} \in \mathbb{R}^{L\times F} $ by function $ \mathcal{E} $, $ \mathcal{M} $, and can be converted to short features $ \tilde{y}_{q,l} \in \mathbb{R}^{L\times T\times F} $ by function $ \mathcal{E} $, $ \mathcal{S} $.
This query embedding $ \tilde{v}_{q,l} $ is then utilized to perform a similarity search across vector databases -- for each layer $ l $, there is a corresponding vector database $ \mathcal{V}_l $ to be searched.
The top $ K $ most similar embeddings $ \left\{\tilde{v}_{k,l}\right\} \in \mathbb{R}^{K\times L\times F} $ are retrieved, along with their associated short features $ \left\{\tilde{y}_{k,l}\right\} \in \mathbb{R}^{K\times L\times T\times F} $.
These $ K $ most relevant audio samples serve as references for detailed comparison with the sample to be detected.
By analyzing the similarities and differences, the authenticity of the tested samples can be better determined.

\subsubsection*{(3) Get Results (Sample Detection)}

As shown in stage 3 (green section) of Figure \ref{fig:ragd}-RAD, our proposed RAD framework requires the training of an additional detection model.
This model accepts the samples to be tested as well as the most relevant retrieved samples.
Specifically, the detection model is provided with the query short features $ \tilde{y}_{q,l} $ and the top $ K $ similar short features $ \left\{\tilde{y}_{k,l}\right\} $ to make the final decision $ z $.
Importantly, this detection model not only evaluates relevant samples, but also provides detailed comparisons with the most similar real samples.
This additional contextual information helps to make more accurate judgments for DF detections.

\subsubsection*{Properties similar to RAG in RAD}

\begin{figure}[t]
  \includegraphics[width=0.475\textwidth]{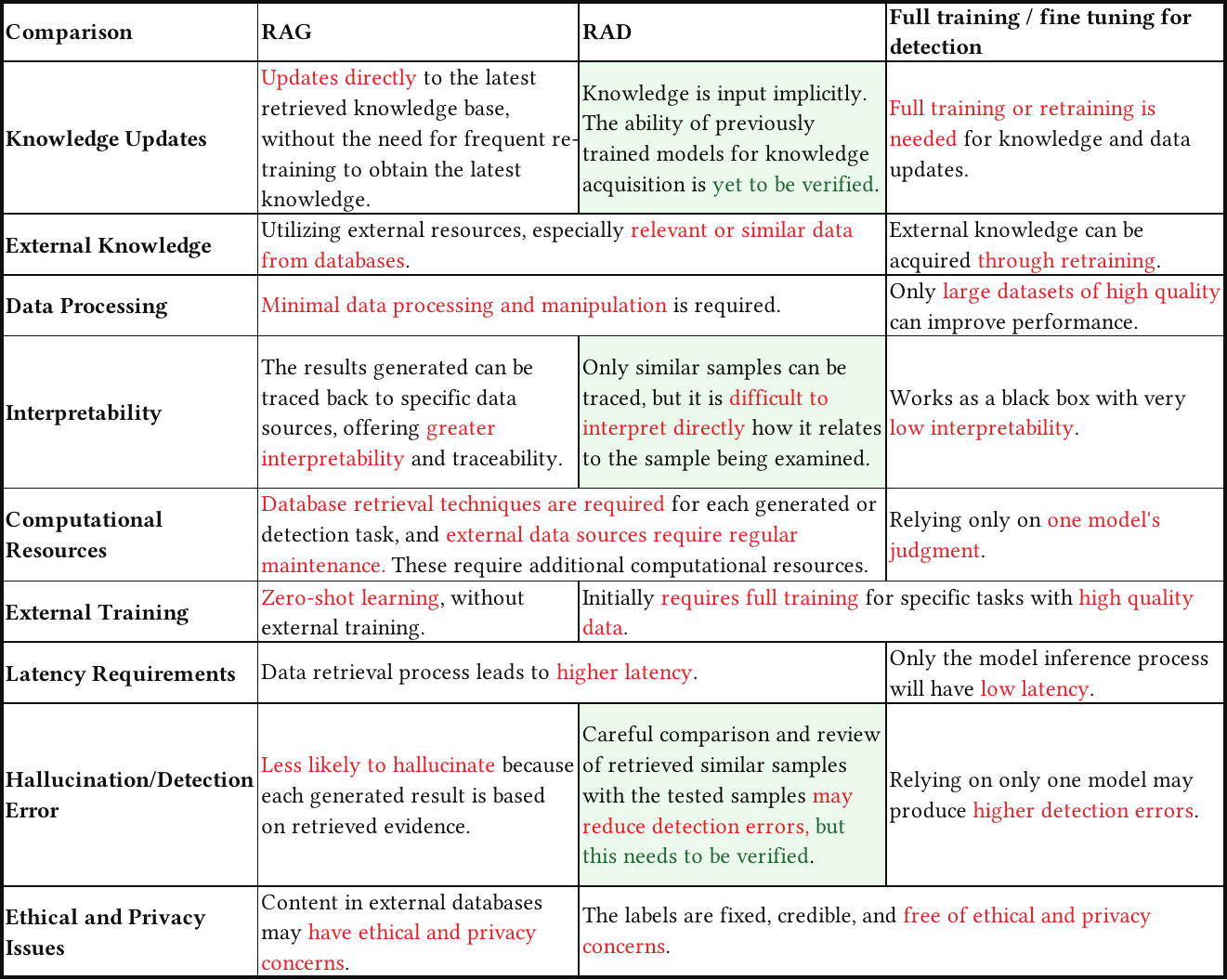}
  \centering
  \caption{Properties of RAG, RAD, and full training/fine-tuning for detection.
    Red text represents the focused attention, and green cells represent ideas that should be verified in this paper.}
  \label{fig:table}
\end{figure}

Despite their different applications, with RAD optimized for detection and RAG for generation, given the similarities in structure and algorithms between RAG and RAD, it is likely that RAD also has the same advantages as RAG.
Figure \ref{fig:table} provides a detailed summary of the key advantages and disadvantages of RAG, RAD, and full training / fine-tuning approaches.
Although similarities and differences exist across these methods, three critical questions emerge as follows:

\begin{itemize}
  \item \textit{\textbf{Question 1:} Does the RAD framework reduce detection errors?}
  \item \textit{\textbf{Question 2:} Does updating external knowledge for the RAD framework further improve detection performance?}
  \item \textit{\textbf{Question 3:} Can the retrieved audio samples be interpreted?}
\end{itemize}

Research questions 1, 2 are verified in $ \S $ \ref{sec:ablation}, and question 3 is verified in $ \S $ \ref{sec:sample}.

\subsection{Detection Model}
\label{sec:detect}

To apply RAD to DF detection, we extend the Multi-Fusion Attentive (MFA) classifier \cite{Guo2023AudioDD}, named RAD-MFA, which combines the raw query input for detection and the retrieved similar bonafide samples to make comprehensive analysis for detections.
Specifically, Figure \ref{fig:radmfa} illustrates the overall structure of our proposed detection model, and Figure \ref{fig:radmfa} shows the MFA sub-modules in detail.

\subsubsection*{MFA Module}

\begin{figure}[t]
  \includegraphics[width=0.475\textwidth]{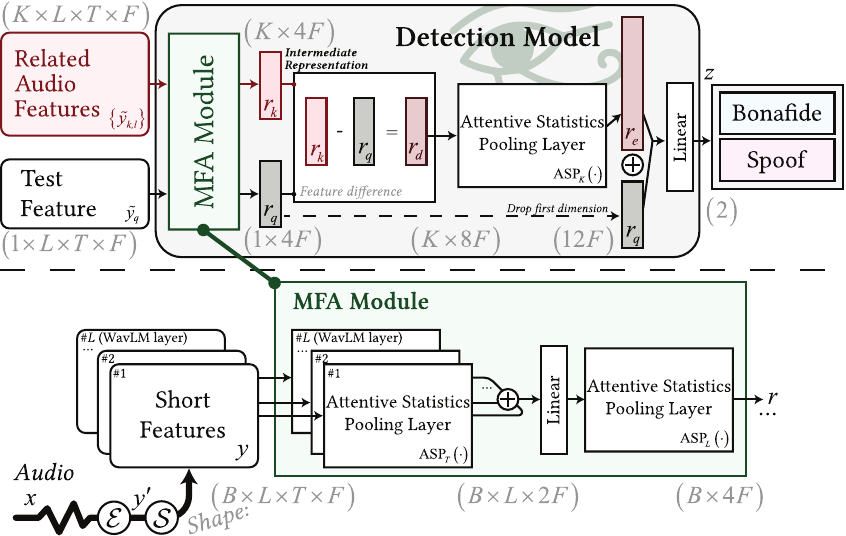}
  \centering
  \caption{The structure of detection model architecture.
    $ \oplus $ denotes the concatenation.
    This process illustrates the $ 3^\mathrm{rd} $ \textit{get results} stage of Figure \ref{fig:ragd}-RAD in detail.}
  \label{fig:radmfa}
\end{figure}

The MFA Module in our framework handles the test feature $ \tilde{y}_q $ and the retrieved features $ \left\{\tilde{y}_{k,l}\right\} $.
For conciseness, these features are denoted by $ y $ in Figure \ref{fig:radmfa}.
Specifically, the MFA module is implemented through the following steps:

\begin{enumerate}[(1)]
  \item The input feature $ y \in \mathbb{R}^{B\times L\times T\times F} $ is passed through $ L $ parallel time-wise attentive statistic pooling (ASP) layers (denoted as $ \mathrm{ASP}_T(\cdot) $) to eliminate the time dimension.
        Here, $ B $ denotes a virtual dimension.
  \item The last outputs are concatenated and passed through a fully connected layer to transform the features to $ \mathbb{R}^{B\times L\times 2F} $.
  \item These outputs are then passed through a layer-wise ASP layer (denoted as $ \mathrm{ASP}_L(\cdot) $) to form the intermediate representation $ r \in \mathbb{R}^{B\times 4F} $.
\end{enumerate}

\subsubsection*{Extended RAD-based MFA}

The RAD-MFA is implemented through the following steps:

\begin{enumerate}[(1)]
  \item The test feature $ \tilde{y}_q \in \mathbb{R}^{1\times L\times T\times F} $ and the retrieved features $ \left\{\tilde{y}_{k,l}\right\} \in \mathbb{R}^{K\times L\times T\times F} $ are sent to the same MFA module, creating intermediate representations $ r_q \in \mathbb{R}^{1\times 4F} $ and $ r_k \in \mathbb{R}^{1\times 4F} $ respectively.
  \item These two representations are used to form $ r_d \in \mathbb{R}^{1\times 4F} $ by taking their difference, $ r_d=r_k-r_q $.
        We make a difference between two features with extremely similar timbre, which allows the discriminative model to pay more attention to other differential information, such as background noise.
  \item This output is sent to a sample-wise ASP layer (denoted as $ \mathrm{ASP}_K(\cdot) $) to form the intermediate representation $ r_e \in \mathbb{R}^{1\times 8F} $.
  \item $ r_e $ is concatenated with $ r_q $, and sent to a fully connected layer to make the final decision.
\end{enumerate}

Through this scheme, the RAD-based detection model can take into account numerous particularly similar bonafide samples and make comprehensive judgments on their contents and distributions.
This enables the model to achieve more accurate detection results by accounting for many additional highly similar authentic cases.

\subsection{Performance Optimization}
\label{sec:performance}

In order to speed up the process of training and testing, two approaches are used for optimization.

\subsubsection*{Locally Stored Features}
\label{sec:store}

To speed up the training and testing process, we pre-compute and cache the WavLM features of all audio segments in the knowledge retrieval database.
Specifically, each raw audio segment $ x_n $ is passed through the WavLM model to extract the corresponding feature representation $ y_{n,l} $.
The retrieved feature $ \left\{y_{n,l}\right\} $ are stored locally, and the indexes in the database only store pointers to these pre-extracted feature paths.
For training, the entire pipeline operates on the pre-computed features rather than raw audio, with the cached test features, retrieved database features, and corresponding labels, making the training process extremely fast.
For testing, this design allows efficient retrieval of audio contents without repetitively invoking the computationally expensive WavLM feature extraction each time.

\subsubsection*{Time-wise Speedup}
\label{sec:speedup}

The acoustic features extracted by WavLM have a frame-to-frame hop size of 20 ms and a large number of layers, resulting in an extremely large number of generated feature parameters that require a large amount of storage space.
To solve the problem, we propose a method $ \mathcal{S}\left(\cdot\right) $ to simplify features in the time dimension.

Let $x$ denote an audio segment sample, the feature extracted by WavLM is called latent long features $ y^\prime \in \mathbb{R}^{T^\prime \times F} $, which can then be transformed into short features $ y \in \mathbb{R}^{T \times F} $ by the $ \mathcal{S} $.
A speedup parameter $ \tau $ is introduced, which partitions the long feature $y^{\prime}$ along the time dimension as:
\begin{equation}
  \mathrm{partition}\left(y^\prime\right)=\mathop {\underbrace{\left[ \left[ y^\prime_1,...,y^\prime_{\tau} \right] ,\left[ y^\prime_{\tau +1},...,y^\prime_{2\tau} \right] ,...,\left[ ...,y^\prime_{T^\prime} \right] \right] }} \limits_{T^\prime/\tau \,\,\mathrm{partitions}}.
  \label{eq:partition}
\end{equation}
The speedup short feature $ y_t $ can then be derived by taking the average along the partitioned time dimension as:
\begin{equation}
  y=\mathcal{S} \left( y^\prime \right) =\left[ \mathrm{mean} \left( \left[ ...,y^\prime_{\tau} \right] \right) ,...,\mathrm{mean} \left( \left[ ...,y^\prime_{T^\prime} \right] \right) \right].
  \label{eq:S}
\end{equation}

Applying this speedup technique enables significant savings in storage space.
However, there is an additional question that needs to be experimentally verified:

\begin{itemize}
  \item \textit{\textbf{Question 4:} Does time-wise speedup method affect downstream DF detection performance?}
\end{itemize}

This question is validated in $ \S $ \ref{sec:ablation}.

\section{Experiments}

The following section describes the datasets and assessment metrics (described in $ \S $ \ref{sec:data}) used for all of the reported experimental work, as well as details of the reproducible implementation (described in $ \S $ \ref{sec:implement}).
The experimental results (described in $ \S $ \ref{sec:results}) will list the evaluation results compared to the existing SOTA.
The ablation studies (described in $ \S $ \ref{sec:ablation}) will then focus on several experiments related to the \textit{\textbf{Four Research Questions}} collected from $ \S $ \ref{sec:rad}, \ref{sec:performance}, and why our proposed RAD framework could be effective.

\subsection{Datasets and Metrics}
\label{sec:data}

\subsubsection*{ASVspoof 2019 LA Database}

The ASVspoof 2019 \cite{Todisco2019ASVspoof2F} logical access (LA) dataset is comprised of bonafide and spoofed utterances generated using totally 19 different spoofing algorithms, including TTS, VC, and replay attacks.
The dataset contains separate partitions for training, development, and evaluation.
The training and development sets contain samples of 6 spoofing algorithms, while the evaluation set contains samples from 2 algorithms seen during training as well as 11 unseen spoofing algorithms not present in the training data.
The training set trains the model, the development set selects the best-performing model, and the evaluation set impartially evaluates the performance of the selected model.
In addition, all bonafide samples will be used to build the retrieval database.

This experimental design aims to evaluate the generalization ability of the DF detection system against unknown spoofing attacks.
Furthermore, the dataset may not contain complete bonafide recordings of all speakers that were impersonated in the spoofing dataset.
The lack of target speaker data may limit the ability of the DF detection system to perform accurate speaker characteristics comparisons.
To mitigate this issue, additional bonafide samples from impersonated speakers should be found to augment the available knowledge database for the retrieval system.

\subsubsection*{ASVspoof 2021 LA Database}

The LA and DF evaluation subsets from the ASVspoof 2021 \cite{Yamagishi2021ASVspoof2A} challenge present intentionally more difficult spoofing detection tasks compared to the 2019 LA data, including more unseen attacks, and both encoding and transmission distortions in the LA set, as well as the unseen coding and compression artifacts in the DF set.
According to the challenge guidelines, since no new training or development data was released for the ASVspoof 2021 challenge, model training, and evaluation were limited only to the ASVspoof 2019 database.
The entire subset of ASVspoof 2021 LA and DF were used for model testing.

\subsubsection*{VCTK Database}

All bonafide samples of ASVSpoof are a very small subset of the VCTK dataset.
To mitigate the issue of insufficient bonafide samples per speaker, as the datasets used originate from the VCTK dataset, we expand our database by using additional samples that exclude all samples already present in our experimental data.
This allows us to increase the number of bonafide samples available for each speaker in the retrieval task.

\subsubsection*{Metrics}

We evaluated our proposed model on these datasets using two standard metrics: the minimum normalized tandem detection cost function (min t-DCF) \cite{Kinnunen2018tDCFAD} and pooled equal error rate (EER).
The min t-DCF measures the combined (tandem) performance of the ASV systems, and the EER reflects the independent DF detection capability.

\subsection{Implementation Details}
\label{sec:implement}

\subsubsection*{Data Processing}

The original audio recordings in the database are segmented into clips of 4 seconds in length.
Audio recordings over the 4-second duration are truncated to 4 seconds.
For audio recordings shorter than 4 seconds, the clips are padded to the 4-second length by repeating the recording.
No additional processing such as voice activity detection (VAD) is applied to the audio samples prior to segmentation.
The audio segments are first encoded into short features using a WavLM model, and are then shortened into more compact short features through a time-wise speedup model with parameter $ \tau=10 $.

\subsubsection*{Vector Database}

In order to store and query embeddings from vector databases more conveniently, we created $ L $ databases to store the audio feature vectors extracted at each layer of the WavLM model.
When performing a database retrieval, the query audio is converted to query embedding, and the top 10 most similar WavLM short features will be retrieved.

\subsubsection*{Model Training}

The front-end feature extractor utilized in this work is WavLM.
During fine-tuning of the front-end WavLM, the Adam optimizer is employed with a learning rate of 3e-6 and a batch size of 4.
For training the MFA, the batch size is changed to 32 and the learning rate is 3e-5.
All experiments were performed utilizing two NVIDIA GeForce RTX 3090 GPUs.
Each model configuration is trained for approximately 30 epochs.

\subsection{Experimental Results}
\label{sec:results}

To demonstrate the superior performance of our proposed method over existing approaches, we compare our proposed method to recent SOTA methods.

\subsubsection*{Results on ASVspoof 2019 LA evaluation set}

The experimental results in Table \ref{tab:asv2019} compare the performance of our proposed methods to existing approaches on the ASVspoof 2019 LA evaluation dataset.
Our method achieves an EER of 0.40\% and a min t-DCF of 0.0115 which is the best reporting result, demonstrating the effectiveness and superiority of our proposed method.
Notably, although Guo et al. \cite{Guo2023AudioDD} utilizes a similar WavLM feature extractor and MFA network, our proposed RAD framework improves its performance, overcoming the limitations of single-model approaches.

In our analysis, the RAD framework first retrieves the most similar audio samples, which are likely from the same speaker, and then performs careful comparisons between these samples and the test sample.
For the detection model, it only needs to consider the differences between the two, rather than relying on fuzzy prior knowledge for detection.
In contrast, our proposed method is more robust.
Specifically, by focusing on fine-grained differences rather than generalized knowledge, our method can more accurately distinguish more detailed information.

\begin{table}[t]
  \setlength{\tabcolsep}{4.9pt}
  \renewcommand{\arraystretch}{1.0}
  \caption{Comparison with other anti-spoofing systems in the ASVspoof 2019 LA evaluation set, reported in terms of pooled min t-DCF and EER (\%).}
  \begin{tabular}{@{}lccc@{}}
    \toprule
    \textbf{System}                               & \textbf{Configuration} & \textbf{min t-DCF} & \textbf{EER(\%)} \\ \midrule
    Hua et al. \cite{Hua2021TowardsES}            & DNN+ResNet             & 0.0481             & 1.64             \\
    Zhang et al. \cite{Zhang2021TheEO}            & FFT+SENet              & 0.0368             & 1.14             \\
    Ding et al. \cite{Ding2022SAMOSA}             & SAMO                   & 0.0356             & 1.08             \\
    Tak et al. \cite{Tak2021EndtoEndSG}           & RawGAT-ST              & 0.0335             & 1.06             \\
    Jung et al. \cite{Jung2021AASISTAA}           & AASIST                 & 0.0275             & 0.83             \\
    Huang et al. \cite{Huang2023DiscriminativeFI} & DFSincNet              & 0.0176             & 0.52             \\
    Fan et al. \cite{Fan2023SpatialRL}            & f0+Res2Net             & 0.0159             & 0.47             \\
    Guo et al. \cite{Guo2023AudioDD}              & WavLM+MFA              & 0.0126             & 0.42             \\ \midrule
    \textbf{Ours}                                 & \textbf{WavLM+RAD-MFA} & \textbf{0.0115}    & \textbf{0.40}    \\ \bottomrule
  \end{tabular}
  \label{tab:asv2019}
\end{table}

\subsubsection*{Results on ASVspoof 2021 DF evaluation set}

We further test our model on the ASVspoof 2021 LA and DF evaluation set, results are shown in Table \ref{tab:asv2021}.
In the DF subset, our method achieves SOTA performance on the DF subset with an EER of 2.38\%.
In the LA subset, we obtain an EER of 4.89\%, which is also quite a competitive performance, but still better than the baseline system \cite{Guo2023AudioDD} without RAD.
Further analysis and ablation studies are needed to a fully characterize the advantages of our proposed method on each component.

\begin{table}[t]
  \setlength{\tabcolsep}{11.5pt}
  \renewcommand{\arraystretch}{1.0}
  \caption{Comparative results of our proposed method with other systems in the ASVspoof 2021 LA and DF evaluation set with pooled EER (\%).}
  \begin{tabular}{@{}lccc@{}}
    \toprule
    \textbf{System}                            & \textbf{Configuration} & \textbf{LA}   & \textbf{DF}   \\ \midrule
    Fan et al. \cite{Fan2023SpatialRL}         & f0+Res2Net             & 3.61          & --            \\
    Do\~nas et al. \cite{MartinDonas2022TheVA} & wav2vec2+ASP           & 3.54          & 4.98          \\
    Wang et al. \cite{Wang2021InvestigatingSF} & wav2vec2+LGF           & 6.53          & 4.75          \\
    Tak et al. \cite{Tak2022AutomaticSV}       & wav2vec2+AASIST        & \textbf{0.82} & 2.85          \\
    Fan et al. \cite{Fan2023SpatialRL}         & WavLM+MFA              & 5.08          & 2.56          \\ \midrule
    \textbf{Ours}                              & \textbf{WavLM+RAD-MFA} & 4.83          & \textbf{2.38} \\ \bottomrule
  \end{tabular}
  \label{tab:asv2021}
\end{table}

\subsection{Ablation Study}
\label{sec:ablation}

\subsubsection*{Ablation Study on Different Components}

\begin{table}[t]
  \setlength{\tabcolsep}{9.6pt}
  \renewcommand{\arraystretch}{1.0}
  \caption{Ablation studies on ASVspoof 2021 DF dataset for the effectiveness of each component with pooled EER (\%).
    -L and -S: large and small.
    ft: fine-tuning.
    Just Difference is the $ r_e $ (denoted in Figure \ref{fig:radmfa}, without $ r_q $) directly connected to the fully connected layer for classification.}
  \begin{tabular}{@{}lcc@{}}
    \toprule
    \textbf{Ablation}       & \textbf{Configuration}               & \textbf{Pooled EER(\%)} \\ \midrule
    Full Framework          & --                                   & \textbf{2.38}           \\ \midrule
    w/o RAD                 & Baseline (Figure \ref{fig:baseline}) & 2.90                    \\
    w/o VCTK                & ASVspoof 2019 only                   & 2.54                    \\
    w/o WavLM-L             & WavLM-S                              & 9.15                    \\ \midrule
    \multirow{2}{*}{w/o ft} & WavLM-S                              & 9.62                    \\
                            & WavLM-L                              & 4.98                    \\ \midrule
    Variation Structure     & Just Difference                      & 2.49                    \\ \bottomrule
  \end{tabular}
  \label{tab:ablation}
\end{table}

The ablation study presented in Table \ref{tab:ablation} summarizes the results obtained by evaluating different configurations and components of the proposed system on the ASVspoof 2021 DF subset.
Specifically, the impact of the RAD framework, WavLM-L feature extractor, fine-tuning of the feature extractor, incorporation of additional VCTK datasets for data retrieval, and the structure of the detection network were analyzed.
The experiments were conducted using a time-wise speedup parameter $ \tau=10 $.
System performance was assessed using the pooled EER expressed as a percentage.
The key observations are summarized as follows (line-by-line explanation from Table \ref{tab:ablation}):

\begin{enumerate}[(1)]
  \item The full system reaches the SOTA performance with a pooled EER of 2.38\%.
  \item Removing the proposed RAD framework for similar sample retrieval increases the pooled EER to 2.90\%.
        This validates the effectiveness of the RAD framework, which answers the research \textit{\textbf{Question 1}}.
        However, this result is slightly higher than that of Guo et al. \cite{Guo2023AudioDD}, which may be due to different parameter settings and time-wise speedup operations.
  \item Excluding the supplementary VCTK dataset slightly increases the pooled EER to 2.64\%, indicating that updating the knowledge with additional related data could improve the detection performance, which answers the research \textit{\textbf{Question 2}}.
  \item Replacing WavLM-L with WavLM-S significantly increases the pooled EER to 9.15\%, highlighting the importance of the feature extractor in the overall framework.
  \item Without fine-tuning, the EER rises drastically to 9.62\% and 4.98\% for WavLM-S and WavLM-L respectively.
        This observation clearly highlights the positive influence of fine-tuning in enhancing DF detection performance, since fine-tuning combines spoofed data instead of using bonafide data alone, thereby improving the discriminatory capability of DF samples.
  \item We also tried the variation structure of removing the $ r_q $ branch and directly connecting $ r_e $ (denoted in Figure \ref{fig:radmfa}) to the classifier slightly increases the pooled EER to 2.49\%, suggesting that not only the difference of the feature, but also the original feature play the role for performance improvement.
\end{enumerate}

\subsubsection*{Effect of Time-wise Speedup Parameter}

\begin{table}[t]
  \setlength{\tabcolsep}{23.6pt}
  \renewcommand{\arraystretch}{1.0}
  \caption{Ablation study of the effect of different time-wise speedup parameter $ \tau $ on DF detection performance of ASVspoof 2021 DF dataset, using pooled EER (\%).}
  \begin{tabular}{@{}lccc@{}}
    \toprule
    \textbf{Speedup} ($ \tau= $) & \textbf{5}    & \textbf{10} & \textbf{20} \\ \midrule
    Original                     & 4.68          & 4.98        & 5.45        \\
    Fine-tune                    & \textbf{2.36} & 2.38        & 2.54        \\ \bottomrule
  \end{tabular}
  \label{tab:speedup}
\end{table}

Table \ref{tab:speedup} examines whether time-wise speedup affects the performance of DF detection.
We tested the original and the fine-tuned WavLM-Large feature extractor on the ASVspoof 2021 DF with $ \tau $ is 5, 10, 20, reporting by the pooled EER.
It is difficult to test under $ \tau < 5 $ due to very high computational costs and storage consumption, which need to be addressed in future work.
Before fine-tuning, the performance varies greatly across $ \tau $: the smaller $ \tau $ is, the better the performance, but the much higher computational cost and storage consumption.
After fine-tuning, the gap narrows, suggesting that optimization can reduce the impact of time-wise speedup operation.
Overall, the time-wise speedup operation will affect the performance, but not too much, which answers the research \textit{\textbf{Question 4}}.
Taking these factors into account, we finally chose $ \tau=10 $.
However, we still need better ways to reduce storage and computation, which is an open problem that needs to be investigated.

\subsection{Sample Analysis}
\label{sec:sample}

\begin{figure}[t]
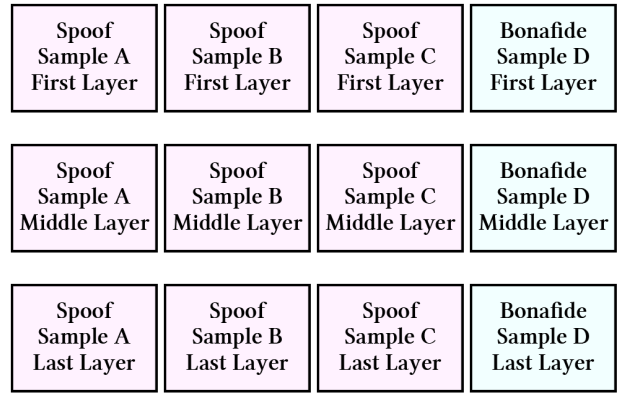

  \makeatletter
  \@for\i:={A0,B0,C0,D0,A1,B1,C1,D1,A2,B2,C2,D2}\do{\FigDigit{\i}}
  \makeatother
  \caption{Examples of retrieved samples \textcolor[rgb]{0.6,0.6,0.6}{(\textit{click on the figure to hear the sound})}.}
  \label{fig:examples}
\end{figure}

The retrieval samples shown in Figure \ref{fig:examples} with clickable audio to hear, offer insights into the factors influencing successful detection of spoofing artifacts.
This figure presents 4 test samples, comprising 3 spoofed and 1 bonafide audio, along with the 2 to 3 most similar samples.
Extracts were taken from three layers (initial, middle, final) of a WavLM-L model.
Despite quality defects in the spoofed test samples, the system appeared to retrieve samples from the same speaker identities.
This suggests the system might rely strongly on speaker-discriminative features, potentially providing an approximate answer to the research \textit{\textbf{Question 3}}.
However, although initial layer retrievals corresponded to the same speakers, middle and final layer results may differ in speaker identity.
Our analysis implies that the shallower layers of the model may focus on timbral and quality-based features, whereas the deeper layers capture more abstract semantic information.
Nevertheless, these explanations remain brief and qualitative, lacking rigorous argumentation, which could be an interesting area for future exploration.
In summary, the retrieved samples have the greatest similarity at the feature level, providing insights for the successful detection of the spoofing artifacts.
After retrieval, a careful comparison of the retrieval results with the test sample will be a key factor in the performance improvement.

\section{Conclusions}

In this work, we proposed a novel retrieval-augmented detection (RAD) framework that leverages retrieved samples to enhance deepfake detection performance.
We also extend the multi-fusion attentive classifier by integrating it with our proposed RAD framework.
Extensive experiments demonstrate state-of-the-art results of the RAD framework on the ASVspoof 2021 DF dataset and competitive performance on the 2019 and 2021 LA datasets.
The consistent improvements achieved on multiple datasets highlight the potential of RAD as a new paradigm for DF detection.
The ablation study reveals RAD and retrieval-augmented generation (RAG) share similar properties in improving detection performance.
Additionally, the retrieved samples are usually from the same speaker, suggesting potential interpretability.
In conclusion, this work opens promising research avenues into retrieval-based augmentation techniques to enhance performance for detection tasks.
By breaking the reliance on a single model, RAD provides a new perspective that utilizes more available information to overcome performance limitations and advance DF detection techniques.

\begin{acks}
  Supported by the Key Research and Development Program of Guangdong Province (grant No. 2021B0101400003) and the Corresponding author is Jianzong Wang (jzwang@188.com).
\end{acks}

\bibliographystyle{ACM-Reference-Format}
\balance
\bibliography{refs}


\begin{thebibliography}{32}


\ifx \showCODEN    \undefined \def \showCODEN     #1{\unskip}     \fi
\ifx \showDOI      \undefined \def \showDOI       #1{#1}\fi
\ifx \showISBNx    \undefined \def \showISBNx     #1{\unskip}     \fi
\ifx \showISBNxiii \undefined \def \showISBNxiii  #1{\unskip}     \fi
\ifx \showISSN     \undefined \def \showISSN      #1{\unskip}     \fi
\ifx \showLCCN     \undefined \def \showLCCN      #1{\unskip}     \fi
\ifx \shownote     \undefined \def \shownote      #1{#1}          \fi
\ifx \showarticletitle \undefined \def \showarticletitle #1{#1}   \fi
\ifx \showURL      \undefined \def \showURL       {\relax}        \fi
\providecommand\bibfield[2]{#2}
\providecommand\bibinfo[2]{#2}
\providecommand\natexlab[1]{#1}
\providecommand\showeprint[2][]{arXiv:#2}

\bibitem[Baevski et~al\mbox{.}(2020)]%
        {Baevski2020wav2vec2A}
\bibfield{author}{\bibinfo{person}{Alexei Baevski}, \bibinfo{person}{Yuhao Zhou}, \bibinfo{person}{Abdelrahman Mohamed}, {and} \bibinfo{person}{Michael Auli}.} \bibinfo{year}{2020}\natexlab{}.
\newblock \showarticletitle{wav2vec 2.0: A framework for self-supervised learning of speech representations}.
\newblock \bibinfo{journal}{\emph{Advances In Neural Information Processing Systems}}  \bibinfo{volume}{33} (\bibinfo{year}{2020}), \bibinfo{pages}{12449--12460}.
\newblock


\bibitem[Cai and Li(2024)]%
        {cai2024integrating}
\bibfield{author}{\bibinfo{person}{Zexin Cai} {and} \bibinfo{person}{Ming Li}.} \bibinfo{year}{2024}\natexlab{}.
\newblock \showarticletitle{Integrating frame-level boundary detection and deepfake detection for locating manipulated regions in partially spoofed audio forgery attacks}.
\newblock \bibinfo{journal}{\emph{Computer Speech \& Language}}  \bibinfo{volume}{85} (\bibinfo{year}{2024}), \bibinfo{pages}{101597}.
\newblock


\bibitem[Chen et~al\mbox{.}(2024)]%
        {chen2024gaia}
\bibfield{author}{\bibinfo{person}{Jinggang Chen}, \bibinfo{person}{Junjie Li}, \bibinfo{person}{Xiaoyang Qu}, \bibinfo{person}{Jianzong Wang}, \bibinfo{person}{Jiguang Wan}, {and} \bibinfo{person}{Jing Xiao}.} \bibinfo{year}{2024}\natexlab{}.
\newblock \showarticletitle{GAIA: Delving into Gradient-based Attribution Abnormality for Out-of-distribution Detection}.
\newblock \bibinfo{journal}{\emph{Advances in Neural Information Processing Systems (NIPS)}}  \bibinfo{volume}{36} (\bibinfo{year}{2024}).
\newblock


\bibitem[Chen et~al\mbox{.}(2021)]%
        {Chen2021WavLMLS}
\bibfield{author}{\bibinfo{person}{Sanyuan Chen}, \bibinfo{person}{Chengyi Wang}, \bibinfo{person}{Zhengyang Chen}, \bibinfo{person}{Yu Wu}, \bibinfo{person}{Shujie Liu}, \bibinfo{person}{Zhuo Chen}, \bibinfo{person}{Jinyu Li}, \bibinfo{person}{Naoyuki Kanda}, \bibinfo{person}{Takuya Yoshioka}, \bibinfo{person}{Xiong Xiao}, \bibinfo{person}{Jian Wu}, \bibinfo{person}{Long Zhou}, \bibinfo{person}{Shuo Ren}, \bibinfo{person}{Yanmin Qian}, \bibinfo{person}{Yao Qian}, \bibinfo{person}{Micheal Zeng}, {and} \bibinfo{person}{Furu Wei}.} \bibinfo{year}{2021}\natexlab{}.
\newblock \showarticletitle{WavLM: Large-Scale Self-Supervised Pre-Training for Full Stack Speech Processing}.
\newblock \bibinfo{journal}{\emph{IEEE Journal of Selected Topics in Signal Processing}}  \bibinfo{volume}{16} (\bibinfo{year}{2021}), \bibinfo{pages}{1505--1518}.
\newblock


\bibitem[Davis and Mermelstein(1980)]%
        {Davis1980ComparisonOP}
\bibfield{author}{\bibinfo{person}{Steven Davis} {and} \bibinfo{person}{Paul Mermelstein}.} \bibinfo{year}{1980}\natexlab{}.
\newblock \showarticletitle{Comparison of parametric representations for monosyllabic word recognition in continuously spoken sentences}.
\newblock \bibinfo{journal}{\emph{IEEE Transactions on Acoustics, Speech, And Signal Processing}} \bibinfo{volume}{28}, \bibinfo{number}{4} (\bibinfo{year}{1980}), \bibinfo{pages}{357--366}.
\newblock


\bibitem[Ding et~al\mbox{.}(2022)]%
        {Ding2022SAMOSA}
\bibfield{author}{\bibinfo{person}{Sivan Ding}, \bibinfo{person}{You Zhang}, {and} \bibinfo{person}{Zhiyao Duan}.} \bibinfo{year}{2022}\natexlab{}.
\newblock \showarticletitle{SAMO: Speaker Attractor Multi-Center One-Class Learning For Voice Anti-Spoofing}.
\newblock \bibinfo{journal}{\emph{International Conference on Acoustics, Speech and Signal Processing (ICASSP)}} (\bibinfo{year}{2022}), \bibinfo{pages}{1--5}.
\newblock


\bibitem[Fan et~al\mbox{.}(2024)]%
        {Fan2023SpatialRL}
\bibfield{author}{\bibinfo{person}{Cunhang Fan}, \bibinfo{person}{Jun Xue}, \bibinfo{person}{Jianhua Tao}, \bibinfo{person}{Jiangyan Yi}, \bibinfo{person}{Chenglong Wang}, \bibinfo{person}{Chengshi Zheng}, {and} \bibinfo{person}{Zhao Lv}.} \bibinfo{year}{2024}\natexlab{}.
\newblock \showarticletitle{Spatial reconstructed local attention Res2Net with F0 subband for fake speech detection}.
\newblock \bibinfo{journal}{\emph{Neural Networks}} (\bibinfo{year}{2024}), \bibinfo{pages}{106320}.
\newblock


\bibitem[Gao et~al\mbox{.}(2023)]%
        {Gao2023RetrievalAugmentedGF}
\bibfield{author}{\bibinfo{person}{Yunfan Gao}, \bibinfo{person}{Yun Xiong}, \bibinfo{person}{Xinyu Gao}, \bibinfo{person}{Kangxiang Jia}, \bibinfo{person}{Jinliu Pan}, \bibinfo{person}{Yuxi Bi}, \bibinfo{person}{Yi Dai}, \bibinfo{person}{Jiawei Sun}, \bibinfo{person}{Qianyu Guo}, \bibinfo{person}{Meng Wang}, {and} \bibinfo{person}{Haofen Wang}.} \bibinfo{year}{2023}\natexlab{}.
\newblock \showarticletitle{Retrieval-Augmented Generation for Large Language Models: A Survey}.
\newblock


\bibitem[Guo et~al\mbox{.}(2024)]%
        {Guo2023AudioDD}
\bibfield{author}{\bibinfo{person}{Yinlin Guo}, \bibinfo{person}{Haofan Huang}, \bibinfo{person}{Xi Chen}, \bibinfo{person}{He Zhao}, {and} \bibinfo{person}{Yuehai Wang}.} \bibinfo{year}{2024}\natexlab{}.
\newblock \showarticletitle{Audio Deepfake Detection With Self-Supervised Wavlm And Multi-Fusion Attentive Classifier}. In \bibinfo{booktitle}{\emph{International Conference on Acoustics, Speech and Signal Processing (ICASSP)}}. IEEE, \bibinfo{pages}{12702--12706}.
\newblock


\bibitem[Hamza et~al\mbox{.}(2022)]%
        {hamza2022deepfake}
\bibfield{author}{\bibinfo{person}{Ameer Hamza}, \bibinfo{person}{Abdul Rehman~Rehman Javed}, \bibinfo{person}{Farkhund Iqbal}, \bibinfo{person}{Natalia Kryvinska}, \bibinfo{person}{Ahmad~S Almadhor}, \bibinfo{person}{Zunera Jalil}, {and} \bibinfo{person}{Rouba Borghol}.} \bibinfo{year}{2022}\natexlab{}.
\newblock \showarticletitle{Deepfake audio detection via MFCC features using machine learning}.
\newblock \bibinfo{journal}{\emph{IEEE Access}}  \bibinfo{volume}{10} (\bibinfo{year}{2022}), \bibinfo{pages}{134018--134028}.
\newblock


\bibitem[Hua et~al\mbox{.}(2021)]%
        {Hua2021TowardsES}
\bibfield{author}{\bibinfo{person}{Guang Hua}, \bibinfo{person}{A. Teoh}, {and} \bibinfo{person}{Haijian Zhang}.} \bibinfo{year}{2021}\natexlab{}.
\newblock \showarticletitle{Towards End-to-End Synthetic Speech Detection}.
\newblock \bibinfo{journal}{\emph{IEEE Signal Processing Letters}}  \bibinfo{volume}{28} (\bibinfo{year}{2021}), \bibinfo{pages}{1265--1269}.
\newblock


\bibitem[Huang et~al\mbox{.}(2023)]%
        {Huang2023DiscriminativeFI}
\bibfield{author}{\bibinfo{person}{Bingyuan Huang}, \bibinfo{person}{Sanshuai Cui}, \bibinfo{person}{Jiwu Huang}, {and} \bibinfo{person}{Xiangui Kang}.} \bibinfo{year}{2023}\natexlab{}.
\newblock \showarticletitle{Discriminative Frequency Information Learning for End-to-End Speech Anti-Spoofing}.
\newblock \bibinfo{journal}{\emph{IEEE Signal Processing Letters}}  \bibinfo{volume}{30} (\bibinfo{year}{2023}), \bibinfo{pages}{185--189}.
\newblock


\bibitem[Kawa et~al\mbox{.}(2023)]%
        {Kawa2023ImprovedDD}
\bibfield{author}{\bibinfo{person}{Piotr Kawa}, \bibinfo{person}{Marcin Plata}, \bibinfo{person}{Michal Czuba}, \bibinfo{person}{Piotr Szyma'nski}, {and} \bibinfo{person}{Piotr Syga}.} \bibinfo{year}{2023}\natexlab{}.
\newblock \showarticletitle{Improved DeepFake Detection Using Whisper Features}.
\newblock \bibinfo{journal}{\emph{International Speech Communication Association (Interspeech)}}  \bibinfo{volume}{abs/2306.01428} (\bibinfo{year}{2023}).
\newblock


\bibitem[Kinnunen et~al\mbox{.}(2018)]%
        {Kinnunen2018tDCFAD}
\bibfield{author}{\bibinfo{person}{Tomi~H. Kinnunen}, \bibinfo{person}{Kong-Aik Lee}, \bibinfo{person}{H{\'e}ctor Delgado}, \bibinfo{person}{Nicholas W.~D. Evans}, \bibinfo{person}{Massimiliano Todisco}, \bibinfo{person}{Md. Sahidullah}, \bibinfo{person}{Junichi Yamagishi}, {and} \bibinfo{person}{Douglas~A. Reynolds}.} \bibinfo{year}{2018}\natexlab{}.
\newblock \showarticletitle{t-DCF: a Detection Cost Function for the Tandem Assessment of Spoofing Countermeasures and Automatic Speaker Verification}.
\newblock  (\bibinfo{year}{2018}).
\newblock


\bibitem[Lewis et~al\mbox{.}(2020)]%
        {Lewis2020RetrievalAugmentedGF}
\bibfield{author}{\bibinfo{person}{Patrick Lewis}, \bibinfo{person}{Ethan Perez}, \bibinfo{person}{Aleksandra Piktus}, \bibinfo{person}{Fabio Petroni}, \bibinfo{person}{Vladimir Karpukhin}, \bibinfo{person}{Naman Goyal}, \bibinfo{person}{Heinrich K{\"u}ttler}, \bibinfo{person}{Mike Lewis}, \bibinfo{person}{Wen-tau Yih}, \bibinfo{person}{Tim Rockt{\"a}schel}, {et~al\mbox{.}}} \bibinfo{year}{2020}\natexlab{}.
\newblock \showarticletitle{Retrieval-augmented generation for knowledge-intensive NLP tasks}.
\newblock \bibinfo{journal}{\emph{Advances in Neural Information Processing Systems}}  \bibinfo{volume}{33} (\bibinfo{year}{2020}), \bibinfo{pages}{9459--9474}.
\newblock


\bibitem[Li et~al\mbox{.}(2021)]%
        {li2021channel}
\bibfield{author}{\bibinfo{person}{Xu Li}, \bibinfo{person}{Xixin Wu}, \bibinfo{person}{Hui Lu}, \bibinfo{person}{Xunying Liu}, {and} \bibinfo{person}{Helen Meng}.} \bibinfo{year}{2021}\natexlab{}.
\newblock \showarticletitle{Channel-wise gated res2net: Towards robust detection of synthetic speech attacks}.
\newblock \bibinfo{journal}{\emph{International Speech Communication Association (Interspeech)}} (\bibinfo{year}{2021}).
\newblock


\bibitem[Luo et~al\mbox{.}(2021)]%
        {luo2021capsule}
\bibfield{author}{\bibinfo{person}{Anwei Luo}, \bibinfo{person}{Enlei Li}, \bibinfo{person}{Yongliang Liu}, \bibinfo{person}{Xiangui Kang}, {and} \bibinfo{person}{Z~Jane Wang}.} \bibinfo{year}{2021}\natexlab{}.
\newblock \showarticletitle{A capsule network based approach for detection of audio spoofing attacks}. In \bibinfo{booktitle}{\emph{International Conference on Acoustics, Speech and Signal Processing (ICASSP)}}. IEEE, \bibinfo{pages}{6359--6363}.
\newblock


\bibitem[Mart'in-Donas and {\'A}lvarez(2022)]%
        {MartinDonas2022TheVA}
\bibfield{author}{\bibinfo{person}{Juan~M. Mart'in-Donas} {and} \bibinfo{person}{Aitor {\'A}lvarez}.} \bibinfo{year}{2022}\natexlab{}.
\newblock \showarticletitle{The Vicomtech Audio Deepfake Detection System Based on Wav2vec2 for the 2022 ADD Challenge}.
\newblock \bibinfo{journal}{\emph{International Conference on Acoustics, Speech and Signal Processing (ICASSP)}} (\bibinfo{year}{2022}), \bibinfo{pages}{9241--9245}.
\newblock


\bibitem[Radford et~al\mbox{.}(2023)]%
        {Radford2022RobustSR}
\bibfield{author}{\bibinfo{person}{Alec Radford}, \bibinfo{person}{Jong~Wook Kim}, \bibinfo{person}{Tao Xu}, \bibinfo{person}{Greg Brockman}, \bibinfo{person}{Christine McLeavey}, {and} \bibinfo{person}{Ilya Sutskever}.} \bibinfo{year}{2023}\natexlab{}.
\newblock \showarticletitle{Robust speech recognition via large-scale weak supervision}.
\newblock  (\bibinfo{year}{2023}), \bibinfo{pages}{28492--28518}.
\newblock


\bibitem[Sun et~al\mbox{.}(2023)]%
        {Sun2023AISynthesizedVD}
\bibfield{author}{\bibinfo{person}{Chengzhe Sun}, \bibinfo{person}{Shan Jia}, \bibinfo{person}{Shuwei Hou}, {and} \bibinfo{person}{Siwei Lyu}.} \bibinfo{year}{2023}\natexlab{}.
\newblock \showarticletitle{AI-Synthesized Voice Detection Using Neural Vocoder Artifacts}.
\newblock \bibinfo{journal}{\emph{2023 IEEE/CVF Conference on Computer Vision and Pattern Recognition Workshops (CVPRW)}} (\bibinfo{year}{2023}), \bibinfo{pages}{904--912}.
\newblock


\bibitem[Tak et~al\mbox{.}(2022)]%
        {Tak2022AutomaticSV}
\bibfield{author}{\bibinfo{person}{Hemlata Tak}, \bibinfo{person}{Massimiliano Todisco}, \bibinfo{person}{Xin Wang}, \bibinfo{person}{Jee weon Jung}, \bibinfo{person}{Junichi Yamagishi}, {and} \bibinfo{person}{Nicholas W.~D. Evans}.} \bibinfo{year}{2022}\natexlab{}.
\newblock \showarticletitle{Automatic speaker verification spoofing and deepfake detection using wav2vec 2.0 and data augmentation}.
\newblock \bibinfo{journal}{\emph{Speaker Odyssey Workshop}} (\bibinfo{year}{2022}).
\newblock


\bibitem[Tak et~al\mbox{.}(2021)]%
        {Tak2021EndtoEndSG}
\bibfield{author}{\bibinfo{person}{Hemlata Tak}, \bibinfo{person}{Jee weon Jung}, \bibinfo{person}{Jose Patino}, \bibinfo{person}{Madhu~R. Kamble}, \bibinfo{person}{Massimiliano Todisco}, {and} \bibinfo{person}{Nicholas W.~D. Evans}.} \bibinfo{year}{2021}\natexlab{}.
\newblock \showarticletitle{End-to-End Spectro-Temporal Graph Attention Networks for Speaker Verification Anti-Spoofing and Speech Deepfake Detection}.
\newblock \bibinfo{journal}{\emph{ASVspoof 2021 Workshop-Automatic Speaker Verification and Spoofing Coutermeasures Challenge}} (\bibinfo{year}{2021}).
\newblock


\bibitem[Todisco et~al\mbox{.}(2019)]%
        {Todisco2019ASVspoof2F}
\bibfield{author}{\bibinfo{person}{Massimiliano Todisco}, \bibinfo{person}{Xin Wang}, \bibinfo{person}{Ville Vestman}, \bibinfo{person}{Md. Sahidullah}, \bibinfo{person}{H{\'e}ctor Delgado}, \bibinfo{person}{Andreas Nautsch}, \bibinfo{person}{Junichi Yamagishi}, \bibinfo{person}{Nicholas W.~D. Evans}, \bibinfo{person}{Tomi~H. Kinnunen}, {and} \bibinfo{person}{Kong-Aik Lee}.} \bibinfo{year}{2019}\natexlab{}.
\newblock \showarticletitle{ASVspoof 2019: Future Horizons in Spoofed and Fake Audio Detection}. In \bibinfo{booktitle}{\emph{International Speech Communication Association (Interspeech)}}.
\newblock


\bibitem[Wang and Yamagishi(2021)]%
        {wang2021comparative}
\bibfield{author}{\bibinfo{person}{Xin Wang} {and} \bibinfo{person}{Junich Yamagishi}.} \bibinfo{year}{2021}\natexlab{}.
\newblock \showarticletitle{A comparative study on recent neural spoofing countermeasures for synthetic speech detection}.
\newblock \bibinfo{journal}{\emph{International Speech Communication Association (Interspeech)}} (\bibinfo{year}{2021}).
\newblock


\bibitem[Wang and Yamagishi(2022)]%
        {Wang2021InvestigatingSF}
\bibfield{author}{\bibinfo{person}{Xin Wang} {and} \bibinfo{person}{Junichi Yamagishi}.} \bibinfo{year}{2022}\natexlab{}.
\newblock \showarticletitle{Investigating self-supervised front ends for speech spoofing countermeasures}.
\newblock \bibinfo{journal}{\emph{The Speaker and Language Recognition Workshop}}  \bibinfo{volume}{abs/2111.07725} (\bibinfo{year}{2022}).
\newblock


\bibitem[weon Jung et~al\mbox{.}(2021)]%
        {Jung2021AASISTAA}
\bibfield{author}{\bibinfo{person}{Jee weon Jung}, \bibinfo{person}{Hee-Soo Heo}, \bibinfo{person}{Hemlata Tak}, \bibinfo{person}{Hye jin Shim}, \bibinfo{person}{Joon~Son Chung}, \bibinfo{person}{Bong-Jin Lee}, \bibinfo{person}{Ha jin Yu}, {and} \bibinfo{person}{Nicholas W.~D. Evans}.} \bibinfo{year}{2021}\natexlab{}.
\newblock \showarticletitle{AASIST: Audio Anti-Spoofing Using Integrated Spectro-Temporal Graph Attention Networks}.
\newblock \bibinfo{journal}{\emph{International Conference on Acoustics, Speech and Signal Processing (ICASSP)}} (\bibinfo{year}{2021}), \bibinfo{pages}{6367--6371}.
\newblock


\bibitem[Yamagishi et~al\mbox{.}(2021)]%
        {Yamagishi2021ASVspoof2A}
\bibfield{author}{\bibinfo{person}{Junichi Yamagishi}, \bibinfo{person}{Xin Wang}, \bibinfo{person}{Massimiliano Todisco}, \bibinfo{person}{Md Sahidullah}, \bibinfo{person}{Jose Patino}, \bibinfo{person}{Andreas Nautsch}, \bibinfo{person}{Xuechen Liu}, \bibinfo{person}{Kong~Aik Lee}, \bibinfo{person}{Tomi Kinnunen}, \bibinfo{person}{Nicholas Evans}, {et~al\mbox{.}}} \bibinfo{year}{2021}\natexlab{}.
\newblock \showarticletitle{ASVspoof 2021: accelerating progress in spoofed and deepfake speech detection}. In \bibinfo{booktitle}{\emph{ASVspoof 2021 Workshop-Automatic Speaker Verification and Spoofing Coutermeasures Challenge}}.
\newblock


\bibitem[Yi et~al\mbox{.}(2022)]%
        {yi2022add}
\bibfield{author}{\bibinfo{person}{Jiangyan Yi}, \bibinfo{person}{Ruibo Fu}, \bibinfo{person}{Jianhua Tao}, \bibinfo{person}{Shuai Nie}, \bibinfo{person}{Haoxin Ma}, \bibinfo{person}{Chenglong Wang}, \bibinfo{person}{Tao Wang}, \bibinfo{person}{Zhengkun Tian}, \bibinfo{person}{Ye Bai}, \bibinfo{person}{Cunhang Fan}, {et~al\mbox{.}}} \bibinfo{year}{2022}\natexlab{}.
\newblock \showarticletitle{ADD 2022: the first audio deep synthesis detection challenge}. In \bibinfo{booktitle}{\emph{International Conference on Acoustics, Speech and Signal Processing (ICASSP)}}. IEEE, \bibinfo{pages}{9216--9220}.
\newblock


\bibitem[Yi et~al\mbox{.}(2023a)]%
        {Yi2023ADD2T}
\bibfield{author}{\bibinfo{person}{Jiangyan Yi}, \bibinfo{person}{Jianhua Tao}, \bibinfo{person}{Ruibo Fu}, \bibinfo{person}{Xinrui Yan}, \bibinfo{person}{Chenglong Wang}, \bibinfo{person}{Tao Wang}, \bibinfo{person}{Chu~Yuan Zhang}, \bibinfo{person}{Xiaohui Zhang}, \bibinfo{person}{Yan Zhao}, \bibinfo{person}{Yong Ren}, \bibinfo{person}{Leling Xu}, \bibinfo{person}{Jun Zhou}, \bibinfo{person}{Hao Gu}, \bibinfo{person}{Zhengqi Wen}, \bibinfo{person}{Shan Liang}, \bibinfo{person}{Zheng Lian}, \bibinfo{person}{Shuai Nie}, {and} \bibinfo{person}{Haizhou Li}.} \bibinfo{year}{2023}\natexlab{a}.
\newblock \showarticletitle{ADD 2023: the Second Audio Deepfake Detection Challenge}.
\newblock \bibinfo{journal}{\emph{ArXiv}}  \bibinfo{volume}{abs/2305.13774} (\bibinfo{year}{2023}).
\newblock


\bibitem[Yi et~al\mbox{.}(2023b)]%
        {Yi2023AudioDD}
\bibfield{author}{\bibinfo{person}{Jiangyan Yi}, \bibinfo{person}{Chenglong Wang}, \bibinfo{person}{Jianhua Tao}, \bibinfo{person}{Xiaohui Zhang}, \bibinfo{person}{Chu~Yuan Zhang}, {and} \bibinfo{person}{Yan Zhao}.} \bibinfo{year}{2023}\natexlab{b}.
\newblock \showarticletitle{Audio Deepfake Detection: A Survey}.
\newblock \bibinfo{journal}{\emph{ArXiv}}  \bibinfo{volume}{abs/2308.14970} (\bibinfo{year}{2023}).
\newblock


\bibitem[Zhang et~al\mbox{.}(2021)]%
        {Zhang2021TheEO}
\bibfield{author}{\bibinfo{person}{Yuxiang Zhang}, \bibinfo{person}{Wenchao Wang}, {and} \bibinfo{person}{Pengyuan Zhang}.} \bibinfo{year}{2021}\natexlab{}.
\newblock \showarticletitle{The Effect of Silence and Dual-Band Fusion in Anti-Spoofing System}. In \bibinfo{booktitle}{\emph{International Speech Communication Association (Interspeech)}}.
\newblock


\bibitem[Ziabary and Veisi(2021)]%
        {ziabary2021countermeasure}
\bibfield{author}{\bibinfo{person}{Pedram~Abdzadeh Ziabary} {and} \bibinfo{person}{Hadi Veisi}.} \bibinfo{year}{2021}\natexlab{}.
\newblock \showarticletitle{A countermeasure based on cqt spectrogram for deepfake speech detection}. In \bibinfo{booktitle}{\emph{2021 7th International Conference on Signal Processing and Intelligent Systems (ICSPIS)}}. IEEE, \bibinfo{pages}{1--5}.
\newblock


\end{thebibliography}


\end{document}